\documentclass[a4paper]{article}
\usepackage{amssymb}
\usepackage{amsmath}
\usepackage[dvips]{graphicx}
\begin{document}

\begin{center}
{\Large\bf Vector Inflation}

\vspace{1cm}

{\bf Alexey Golovnev,\quad Viatcheslav Mukhanov,\quad Vitaly Vanchurin}
\vspace{0,5cm}

ASC, Department f{\" u}r Physik, Ludwig-Maximilians-Universit{\" a}t M{\" u}nchen,
Theresienstr. 37, D-80333, Munich, Germany
\vspace{0.2cm}

{\small Alexey.Golovnev@physik.uni-muenchen.de;\quad mukhanov@theorie.physik.uni-muenchen.de;\quad vitaly@cosmos.phy.tufts.edu}

\end{center}

\begin{abstract}
We propose a scenario where inflation is driven by non-minimally coupled
massive vector fields. In an isotropic homogeneous universe these fields
behave in presicely the same way as a massive minimally coupled scalar
field. Therefore our model is very similar to the model of chaotic inflation
with scalar field. For vector fields the isotropy of expansion is
achived either by considering a triplet of orthogonal vector fields or for
the expense of $N$ randomly oriented vector fields. In the last case the
substantial anisotropy of the expansion of order $1/\sqrt{N}$ survives until the end of
inflation. The lightest vector fields might also force the late time
acceleration of the Universe.
\end{abstract}

\section{Introduction}

All successful inflationary scenarios are based on the use of classical scalar fields. 
Two main reasons for that are natural homogeneity and
isotropy of such field and its ability to imitate a slowly
decaying cosmological constant. This happens either in models of chaotic inflation \cite{Linde}
or in k-inflation \cite{ADM}. Although
the higher spin bosonic fields can also form condensates they are usually
overlooked since they generically induce an anisotropy and because of the
apparent difficulty to realize the slow-roll regime for them. For example,
a model of vector inflation based on
a potential $V\left( A_{\alpha }A^{\alpha }\right)$ was suggested in Ref. \cite{Ford}, 
where the potential
does not change too much while $I=A_{\alpha }A^{\alpha }$ runs over
exponentially large range during inflation.

In this paper we show that both obstacles for realizing a successful vector
inflation can be simultaneously surmounted in a natural way. In particular, isotropy
of the vector field condensate can be achieved either in the case of a
triplet of mutually orthogonal vector fields \cite{Bertolami,Armendaris} 
(see also \cite{Hoso,Galt} for exact isotropic solutions of the Einstein-Yang-Mills
system based on the same idea) or by
considering a large number of randomly oriented fields.
(Another possibility is to consider purely time-like vector fields \cite{Kis,Sean,Har,Mot}.)
The problem of
slow-roll of the massive vector field can also be successfully solved by
introducing a non-minimal coupling of this field to gravity. As a result,
we obtain inflationary scenarios which are very similar to the simplest
chaotic inflation with a massive scalar field \cite{Linde} and N-flation %
\cite{DKMW}. However, in distinction from N-flation in the case of $N$
vector fields an anisotropy of order $1/\sqrt{N}$ can survive until the end of
inflation.

\section{Equations}

Let us consider a massive vector field, which is non-minimally coupled to
gravity, and has the action%
\begin{equation}
\label{action1}
S=\int dx^{4}\sqrt{-g}\left(-\frac{R}{16\pi }-\frac{1}{4}F_{\mu \nu }F^{\mu \nu }+%
\frac{1}{2}\left( m^{2}+\frac{R}{6}\right) A_{\mu }A^{\mu }\right)
\end{equation}%
where $F_{\mu \nu }\equiv \bigtriangledown _{\mu }A_{\nu }-\bigtriangledown
_{\nu }A_{\mu }=\partial _{\mu }A_{\nu }-\partial _{\nu }A_{\mu }$ and we use
the Planck units ($G=1$). Note that the non-minimal coupling of this vector
field is very similar to conformal coupling for a scalar field. In the
case of a scalar field this coupling converts massless scalar field to
conformal invariant field. As we will see, for the vector field the
particular non-minimal coupling in (\ref{action1}) has
an {}\textquotedblleft opposite effect\textquotedblright , namely, it
violates the conformal invariance of a massless vector field and forces it
to behave in the same way as minimally coupled scalar field. The variation
of the action with respect to $A^{\mu }$ yields the following equations of
motion

\begin{equation*}
\frac{1}{\sqrt{-g}}\frac{\partial }{\partial x^{\mu }}\left( \sqrt{-g}F^{\mu
\nu }\right) +\left( m^{2}+\frac{R}{6}\right) A^{\nu }=0.
\end{equation*}%
In the spatially flat Friedmann universe with the metric 
\begin{equation*}
ds^{2}=dt^{2}-a^{2}(t)\delta _{ik}dx^{i}dx^{k}
\end{equation*}%
these equations take the following form 
\begin{equation}
\label{1}
-\frac{1}{a^{2}}\Delta A_{0}+\left( m^{2}+\frac{R}{6}\right) A_{0}+\frac{1}{a^2%
}\partial _{i}\dot{A}_{i}=0,  
\end{equation}%
\begin{equation}
\label{2}
\ddot{A}_{i}+\frac{\dot{a}}{a}\dot{A}_{i}-\frac{1}{a^{2}}\Delta A_{i}+\left(
m^{2}+\frac{R}{6}\right) A_{i}-\partial _{i}\dot{A}_{0}-\frac{\dot{a}}{a}\partial _{i}{A}_{0}+\frac{1}{%
a^{2}}\partial _{i}\left( \partial _{k}A_{k}\right) =0,  
\end{equation}%
where $\partial _{i}\equiv \partial /\partial x^{i}$, a dot denotes a
derivative with respect to the physical time $t$ and we assume the summation
over repeated spatial indices. One of the quantities which characterize the
strength of the vector field in coordinate independent way is the scalar%
\begin{equation*}
I=A^{\alpha }A_{\alpha }=A_{0}^{2}-\frac{1}{a^{2}}A_{i}A_{i},
\end{equation*}%
and this motivates us to introduce a new variable $B_{i}\equiv A_{i}/a$
instead of $A_{i}$ which is the \textquotedblleft square
root\textquotedblright\ of the invariant $I$. Considering the
quasi-homogeneous vector field ($\partial _{i}A_{\alpha }=0$) we immediately
infer from (\ref{1}) that 
\begin{equation*}
A_{0}=0
\end{equation*}%
and equation (\ref{2}) rewritten in terms of the field strength $B_{i}$
becomes

\begin{equation*}
\ddot{B}_{i}+3H\dot{B}_{i}+m^{2}B_{i}=0,
\end{equation*}%
where $H\equiv \dot{a}/a.$ This equation is similar to the equation for the
massive minimally coupled scalar field, and when the Hubble constant $H$ is
larger than the mass $m$ the fields $B_{i}$ are \textquotedblleft
frozen\textquotedblright . Therefore one might expect that the potential $%
-m^{2}A_{\mu }A^{\mu }=m^{2}B_{i}B_{i}\approx const$ could drive the quasi
de Sitter expansion analogous to the scalar field. To determine under which
conditions this could happen we have to calculate the energy-momentum tensor
for the vector field. Variation of the action (\ref{action1}) with
respect to the metric gives%
\begin{eqnarray}
\label{EMT}
T_{\beta }^{\alpha } &=&\frac{1}{4}F^{\gamma \delta }F_{\gamma \delta
}\delta _{\beta }^{\alpha }-F^{\alpha \gamma }F_{\beta \gamma }+\left( m^{2}+%
\frac{R}{6}\right) A^{\alpha }A_{\beta }-\frac{1}{2}m^{2}A^{\gamma
}A_{\gamma }\delta^{\alpha}_{\beta}  \notag \\
&&+\frac{1}{6}\left( R_{\beta }^{\alpha }-\frac{1}{2}\delta _{\beta
}^{\alpha }R\right) A^{\gamma }A_{\gamma }+\frac{1}{6}\left( \delta _{\beta
}^{\alpha }\square -\nabla ^{\alpha }\nabla _{\beta }\right) A^{\gamma
}A_{\gamma }  
\end{eqnarray}%
For a homogeneous vector field in a flat Friedmann universe we obtain%
\begin{equation}
\label{E1}
T_{0}^{0}=\frac{1}{2}\left( \dot{B}_{k}^{2}+m^{2}B_{k}^{2}\right) ,
\end{equation}%
\begin{eqnarray}
\label{E2}
T_{j}^{i} &=&\left[ -\frac{5}{6}\left( \dot{B}_{k}^{2}-m^{2}B_{k}^{2}\right)
-\frac{2}{3}H\dot{B}_{k}B_{k}-\frac{1}{3}\left( \dot{H}+3H^{2}\right)
B_{k}^{2}\right] \delta _{j}^{i}  \notag \\
&&+\dot{B}_{i}\dot{B}_{j}+H\left( \dot{B}_{i}B_{j}+\dot{B}_{j}B_{i}\right)
+\left( \dot{H}+3H^{2}-m^{2}\right) B_{i}B_{j},
\end{eqnarray}%
where we assume the summation over index $k.$ The spatial part of the
energy-momentum tensor contains off-diagonal components which are of the same
order of magnitude as the diagonal components and therefore the isotropic
Friedmann universe filled by the homogeneous vector field does not satisfy
the Einstein equations. This is not surprising because the homogeneous
vector field has a preferable direction. Therefore to make the
energy-momentum tensor diagonal we have to consider 
several fields simultaneously.

\section{Inflation}

Let us first consider a triplet of mutually orthogonal vector fields $%
B_{i}^{\left( a\right) }$ \cite{Armendaris}, with the same magnitude $\left\vert B\right\vert $
each. Then from 
\begin{equation}
\label{tr1}
\sum\limits_{i}B_{i}^{\left( a\right) }B_{i}^{\left( b\right) }=\left\vert
B\right\vert ^{2}\delta _{b}^{a},
\end{equation}%
it follows that%
\begin{equation*}
\sum\limits_{a}B_{i}^{\left( a\right) }B_{j}^{\left( a\right) }=\left\vert
B\right\vert ^{2}\delta _{j}^{i}.
\end{equation*}%
The last relation is easy to understand considering the components of the
triplet as elements of a matrix; then condition (\ref{tr1}) simply
implies the orthogonality of this matrix. Using these relations we find from
(\ref{E1}), (\ref{E2}) that the total energy-momentum tensor of the vector
fields is%
\begin{equation*}
T_{0}^{0}=\varepsilon =\frac{3}{2}\left( \dot{B}_{k}^{2}+m^{2}B_{k}^{2}%
\right) , 
\end{equation*}%
\begin{equation*}
T_{j}^{i}=-p\delta _{j}^{i}=-\frac{3}{2}\left( \dot{B}%
_{k}^{2}-m^{2}B_{k}^{2}\right)\delta _{j}^{i} 
\end{equation*}%
where $B_{k}$ are the components of any field from the triplet which satisfy%
\begin{equation}
\label{i3}
\ddot{B}_{i}+3H\dot{B}_{i}+m^{2}B_{i}=0,
\end{equation}%
and $H$ is now given by 
\begin{equation*}
H^{2}=4\pi \left( \dot{B}_{k}^{2}+m^{2}B_{k}^{2}\right) .
\end{equation*}%
These equations are precisely the same as for the massive scalar field and
their investigation literally repeats those one for scalar field \cite
{Mukhanovbook}. In particular for $\left\vert B\right\vert >1$ we have
slow-roll regime ($\dot{B}_{k}^{2}\ll m^{2}B_{k}^{2}$) during which $%
p\approx -\varepsilon $ and the universe undergoes the stage of inflation.
The inflation is over when the value of $\left\vert B\right\vert $ drops to
the Planck value. Note that the vector field condensate can also imitate the
ultra-hard equation of state $p=\varepsilon $ when its kinetic energy
dominates.

Another way to resolve the issue of isotropy is to consider a large number
of randomly oriented vector fields. For simplicity, let us first consider $N$
fields with equal masses $m$ assuming that they all have about the same
magnitude of order $B$ initially$.$ The components of these fields satisfy (%
\ref{i3}) and their total contribution to $T_{0}^{0}$ can be estimated as%
\begin{equation*}
T_{0}^{0}=\varepsilon \simeq \frac{N}{2}\left( \dot{B}%
_{k}^{2}+m^{2}B_{k}^{2}\right) .
\end{equation*}%
To get an estimation of the spatial components of the energy-momentum tensor
we note that 
\begin{equation*}
\sum\limits_{a=1}^{N}B_{i}^{\left( a\right) }B_{j}^{\left( a\right) }\simeq 
\frac{N}{3}B^{2}\delta _{j}^{i}+O\left( 1\right) \sqrt{N}B^{2},
\end{equation*}%
where $B^{2}=B_{k}^{2}$ and the summation over $k$ is assumed. The
corrections proportional to $\sqrt{N}$ are due to stochastic random
distribution of directions of the fields and they do not vanish for $%
i\neq j$ characterizing the typical magnitude of the off-diagonal spatial
components of the energy-momentum tensor. It follows from (\ref{E2}) that
during inflation a typical value of the off-diagonal spatial components is
of order $H^{2}\sqrt{N}B^{2}.$ The isotropic inflationary solution is
self-consistent only if these components are smaller than $T_{i}^{i}\sim
T_{0}^{0}\sim H^{2}$; hence this solution can be valid only for $B<1/N^{1/4}.
$ On the other hand, the vector fields are in the slow roll regime only if
the \textquotedblleft effective friction\textquotedblright\ in (\ref{i3}),
which is of order $H$,  exceeds their mass $m$ and inflation is finished when 
$H\simeq m.$ Taking into account that on the inflationary stage 
\begin{equation}
\label{i7}
H^{2}=\frac{8\pi }{3}\varepsilon \simeq \frac{4\pi }{3}Nm^{2}B^{2}
\end{equation}%
we find that when the field drops to $B\simeq 1/N^{1/2}$ it starts to
oscillate and inflation is over. Thus the isotropic inflationary stage of
expansion takes place when the fields $B$ change within the interval%
\begin{equation*}
\frac{1}{\sqrt[4]{N}}>B>\frac{1}{\sqrt{N}}.
\end{equation*}%
To estimate the number of e-folds during the inflationary stage we note that
in the slow-roll regime 
\begin{equation}
 \label{i8}
\dot{B}\approx -\frac{m^{2}B}{3H} 
\end{equation}%
and then from (\ref{i7}), (\ref{i8}) we find that during inflation the scale
factor increases by  
\begin{equation*}
\frac{a_{f}}{a_{i}}\simeq \exp \left( 2\pi NB_{in}^2\right) ,
\end{equation*}%
where $B_{in}$ is the initial value of vector fields. Taking $B_{in}\simeq
N^{-1/4}$ we find that in the case of $N$ vector fields the maximum number
of e-folds of isotropic inflation  is about $2\pi \sqrt{N}$. For $N$ of
order of few hundreds the duration of inflation is enough to explain the
observed homogeneity. On the other hand, at the
end of inflation there still survive the off-diagonal spatial components of
the energy-momentum tensor and their relative value compared to the diagonal
components is about $1/\sqrt{N}.$ They induce a global anisotropy which is
of the same order of magnitude $\simeq 1/\sqrt{N}$ at the end of inflation.
In the case of few hundreds vector fields this anisotropy is about few
percents. 

The consideration above can easily be generalized to the case of many vector
fields with different masses and initial amplitudes. The very light fields
can remain frozen until today and thus serve as the observed dark energy
which can be even anisotropic. 

Moreover, instead of the mass term $%
m^{2}A_{\alpha }A^{\alpha }$ we can consider an arbitrary potential $V\left(
A_{\alpha }A^{\alpha }\right) .$ In this case the vector fields satisfy the
equation
\begin{equation*}
\ddot{B_{i}}+3\frac{\dot{a}}{a}\dot{B}_{i}+V^{\prime }(B^{2})B_{i}=0,
\end{equation*}%
and it is clear that the slow-roll regime can be realized for the
wide class of potentials thus providing us with a large variety of
inflationary scenarios similar to the case of the scalar field. 
The energy-momentum tensor is:
\begin{equation*}
T_{0}^{0}=\frac{1}{2}\left( \dot{B}_{k}^{2}+V\left(B^{2}\right)\right) ,
\end{equation*}%
\begin{eqnarray*}
T_{j}^{i} &=&\left[ -\frac{5}{6}\dot{B}_{k}^{2}+\frac{1}{2}V\left(B^{2}\right)
-\frac{2}{3}H\dot{B}_{k}B_{k}-\frac{1}{3}\left( \dot{H}+3H^{2}-V^{\prime}\left(B^{2}\right)\right)
B_{k}^{2}\right] \delta _{j}^{i}  \notag \\
&&+\dot{B}_{i}\dot{B}_{j}+H\left( \dot{B}_{i}B_{j}+\dot{B}_{j}B_{i}\right)
+\left( \dot{H}+3H^{2}-V^{\prime}\left(B^{2}\right)\right) B_{i}B_{j}
\end{eqnarray*}%
and after averaging over $N$ fields we obtain
\begin{equation*}
T_{j}^{i}=-p\delta^{i}_{j}\simeq \frac{N}{2}\left( -\dot{B}%
_{k}^{2}+V\left(B^{2}\right)\right)\delta^{i}_{j} .
\end{equation*}
For a general potential $V$ the exit from inflation occurs when
$V^{\prime}B/V\sim\sqrt{N}$.
Note that it allows us to make the
anisotropy
smaller because the inflation may end at small
$B_f$ and the corresponding anisotropy is then
of order $\sqrt{N}B^{2}_f$.

\section{Conclusions}

Scalar field inflation predicts nearly completely isotropic universe. The
anisotropy can be obtained only for the expense of initial conditions and
fine tuning the duration of inflationary stage. In this paper, we have
proposed a vector inflation, which either could give us completely isotropic
universe (with orthogonal triplet of vector fields) or slightly anisotropic
universe (with $N$ randomly oriented vector fields). The degree of
anisotropy is of order $1/\sqrt{N}$ at the end of inflation. 

The model contains only minimal complexity and does not require any fine-tuning
of the potential or initial conditions. To implement inflationary
expansion with vector fields, two essential ingredients were added to the
standard theory of massive vector fields. First, to obtain the slow roll
regime for the vector fields we have coupled them to gravity in a non-minimal
way. Second, to avoid a large anisotropy we have considered a large number
of mutually uncoupled randomly oriented fields.

The same model can in principle explain the late time acceleration of the
Universe. In fact, this model naturally combines inflation and dark energy
within the same framework. However, predictions of the unified model
depend crucially on the distribution of masses of vector fields. \newline

\emph{To Andrei Linde on the occasion of his 60th birthday.}%

The authors are grateful to TRR 33 {}\textquotedblleft The Dark
Universe\textquotedblright\ and the Cluster of Excellence EXC 153
{}\textquotedblleft Origin and Structure of the Universe\textquotedblright\
for support.

\end{document}